\documentclass[12pt]{article}
\usepackage{amsmath}
\usepackage{graphicx}
\usepackage{float}
\newcommand{\bb}{\begin{equation}}
\newcommand{\ee}{\end{equation}}
\newcommand{\ba}{\begin{eqnarray}}
\newcommand{\ea}{\end{eqnarray}}

\begin{document}

\title{{\bf Crudely Estimated Maximum Mass for a Cosmological Black Hole}}
%\thanks{Alberta-Thy-24-16, arXiv:xxx.yyyyy [hep-th]}

\author{
Don N. Page
\thanks{Internet address:
profdonpage@gmail.com}
\\
Theoretical Physics Institute\\
Department of Physics\\
4-183 CCIS\\
University of Alberta\\
Edmonton, Alberta T6G 2E1\\
Canada
}

\date{2026 August 25}

\maketitle
\large
\begin{abstract}
\baselineskip 23 pt

$N$-body simulations by M.~Milosavljevic and D.~Merritt \cite{Milosavljevic:2001vi} suggest that for supermassive black hole binaries in their expected nearly circular orbits, before gravitational wave emission dominates, the inspiral rate causes the dimensionless orbital relative velocity squared, $(v/c)^2$, to increase nearly linearly with time at a rate that is approximately the inverse of 210 Gyr (about 15 times the present age of the universe), nearly independently of the mass of the binary.  This by itself would not lead to coalescence by the present time, but when it is augmented by gravitational radiation, it can lead to coalescence for equal-mass black hole binaries up to about $1.4\times 10^{14} M_\odot$, which might then be conjectured to be a very crude upper limit to the mass of black holes in our universe at the present time.  This also suggests that the gravitational wave strain from coalescing supermassive black hole binaries should decrease for wave periods greater than about 100\,000 years.

\end{abstract}

\normalsize

\baselineskip 21.6 pt

\newpage

%\section{Introduction}

The `final-parsec' problem \cite{Begelman:1980vb, Milosavljevic:2001vi, Milosavljevic:2002ht, Lodato:2009qd, Colpi:2009bk, Nixon:2011tn, Khan:2013wbx, Vasiliev:2013nha, Merritt:2013, Vasiliev:2014poa, Aly:2015vqa, Vasiliev:2015uzo, Goicovic:2016dul, Bortolas:2016xpy, Dosopoulou:2016hbg, Ryu:2018yhv, Generozov:2018niv, Lima:2020fzk, Mapelli:2021taw, Spera:2022byb, Mayer:2023tbm, Zhu:2023imz, Koo:2023gfm, Zhang:2023jrk, Alonso-Alvarez:2024gdz, Shi:2024skj, Tiruvaskar:2025lkq, Alexander:2025rtn}
shows that black hole binary evolution purely by the emission of gravitational radiation is not sufficient to bring nearly enough black hole pairs close enough together to match their observation by the gravitational radiation they emit.  The sources cited above give a number of solutions in which black hole pairs are brought close enough together for gravitational radiation to dominate by losing energy to stars or gas.

The $N$-body simulations by Milo\v{s} Milosavljevi\'{c} and David Merritt \cite{Milosavljevic:2001vi} followed the evolution of stars and black hole binaries in galactic nuclei and found that the gravitational effect of nearby stars (but initially here not including the effect of gravitational radiation emitted by the binary) causes the inverse of the semimajor axis $a$ of a black hole binary to increase nearly linearly with time at a rate that they give in their Eqs. (42) and (43) as
\bb
S \equiv \frac{d}{dt}\left(\frac{1}{a}\right) \approx 1.0\times 10^{-6}
\left(\frac{M_{\mathrm{BH}}}{10^8 M_\odot}\right)^{-0.95}
\mathrm{yr}^{-1}\mathrm{pc}^{-1},
\label{S}
\ee
correcting an obvious typo in the last denominator of the original Eq.\ (43).

For simplicity, I shall approximate the exponent $-0.95$ as $-1$ and shall rewrite the time derivative as that of the square of the normalized orbital velocity,
\bb
\left(\frac{v}{c}\right)^2 = \frac{GM}{c^2 a}
\label{v}
\ee

Then Eq.\ (43) of \cite{Milosavljevic:2001vi} becomes approximately
\bb
\frac{d}{dt}\left(\frac{v}{c}\right)^2 \approx \frac{1}{T}
\label{vdot}
\ee
with the timescale (approximately independent of the binary black hole mass when gravitational radiation is omitted) of
\bb
T = \frac{\mathrm{yr}\ \mathrm{pc}}{100\,GM_\odot/c^2} \approx 210\, \mathrm{Gyr} \approx 15\, t_0,
\label{T}
\ee
where $t_0 \approx 14\, \mathrm{Gyr}$ is the current age of the universe.

If one now includes the energy lost to gravitational radiation as calculated by Peters and Mathews 
\cite{Peters:1963ux, Peters:1964qza, Peters:1964zz},
one gets, for black hole masses $M_1$ and $M_2$, total mass $M = M_1 + M_2$, and scaled dimensionless symmetric mass ratio $4\eta = 4M_1M_2/(M_1+M_2)^2 \leq 1$,
\bb
T\frac{d}{dt}\left(\frac{v}{c}\right)^2 \approx 
1 + \frac{16(4\eta)c^3}{5\,GM}T\left(\frac{v}{c}\right)^{10}
\equiv 1 + x^5,
\label{vdotgr}
\ee
where
\bb
x \equiv \left(\frac{16\,(4\eta)\,c^3T}{5\,GM}\right)^{1/5}
\left(\frac{v}{c}\right)^{2} 
\equiv \left(\frac{4\eta\, M_*}{M}\right)^{1/5}
\left(\frac{v}{c}\right)^2,
\label{z}
\ee
with
\bb
M_* \equiv \frac{16}{5}\frac{c^3 T}{G}  
= \frac{4}{125}\frac{c^5\ \mathrm{yr}\ \mathrm{pc}}{(GM_\odot)^2} M_\odot
\approx 4.3\times 10^{24} M_\odot \approx 200\,M_N,
\label{M*}
\ee
where
\bb
M_N = \frac{c^2}{3G\sqrt{\Lambda}} = 2.164(35)\times 10^{22} M_\odot
\label{MN}
\ee
is the maximum mass of a black hole \cite{Ashtekar:2019khv} in a universe with a cosmological constant $\Lambda = 1.088(30)\times 10^{-2}$ m$^{-2}$ \cite{PDG} (or $\Lambda c^2 = 0.978(27)$ times the more memorable approximation of ten square attohertz \cite{Scott:2013oib}).  $M_N$ might be called the Nariai mass, since it is the mass of the black hole in the Nariai metric \cite{Nariai}, which is the limiting case of the Schwarzschild-de Sitter metric in which both the black hole horizon and the cosmological horizon have the same area.
Then if we define
\bb
y \equiv \left(\frac{4\eta\, M_*}{M}\right)^{1/5}\frac{t}{T}
\equiv \left(\frac{4\eta\, M_\odot}{M}\right)^{1/5}\frac{t}{T_*}
\label{y}
\ee
where
\bb
T_* \equiv \left(\frac{M_\odot}{M_*}\right)^{1/5}T
\approx 1.2\times 10^{-5}\, T \approx 2.5\, \mathrm{Myr},
\label{T*}
\ee
and if we assume that $x << 1$ at $t = 0$ and hence at $y = 0$, then
\bb
y(x) = \int_0^x \frac{dz}{1+z^5}.
\label{y(x)}
\ee

Relative to the entire inspiral time, the black holes coalesce very shortly after $(v/c)^2$ becomes of the order of unity and the nonrelativistic approximations that lead to the equations above break down.  However, since $M \leq M_N \ll M_*$,\\ $x$ becomes much larger than unity before $(v/c)^2$ becomes of the order of unity, and the nonrelativistic approximation gives a good estimate for the inspiral time.  Indeed, near the end of the inspiral, when $(v/c)^2 = O(1)$, we have $x \gg 1$,  so then
\bb
y(x) \approx y_\infty = \int_0^\infty \frac{dz}{1+z^5} = \frac{\pi/5}{\sin{\pi/5}} \approx 1.069.
\label{y(infty)}
\ee

Therefore, since $M \leq M_N \ll M_*$, the time needed to go from $(v/c)^2 \ll 1$ to coalescence (shortly after $(v/c)^2$ becomes of the order of unity) is approximately
\bb
t_c \approx T_* y_\infty\left(\frac{M}{4\eta\, M_\odot}\right)^{1/5}
\approx (2.6\,\mathrm{Myr})\left(\frac{M}{4\eta\, M_\odot}\right)^{1/5}.
\label{tap}
\ee
For this time to be less than $t_0$, the current age of the universe, one needs
\bb
M < 4\eta\, M_\odot (t_0/t_c)^5 \approx 3.9\times 10^{18} (4\eta) M_\odot.
\label{Mm}
\ee

However, the requirement for a black hole of huge mass to form is likely to be more severe, since it presumably requires a sequence of binary black hole coalescences.  For simplicity, define the dimensionless binary black hole mass
\bb
m \equiv \frac{M}{4\eta\, M_\odot} \equiv \frac{(M_1+M_2)^3}{4M_\odot M_1M_2} \geq \frac{M}{M_\odot}.
\label{m}
\ee
Suppose the initial dimensionless mass is $m_0$, and that at each of $N$ stages, two equal mass black holes form a black hole of essentially twice the mass, so at the $n$th stage one gets to $m_n = 2^n m_0$ in a time $t_n = T_* y_\infty m_0^{1/5} 2^{n/2}$ for that stage.  For $N \gg 1$ stages, the total time needed is
\bb
t_{\mathrm{total}} = T_* y_\infty m_0^{1/5} \frac{2^{N/5}-1}{1-2^{-1/5}}
\approx \frac{T_* y_\infty}{1-2^{-1/5}}m_0^{1/5}2^{N/5}
\approx (2.0\times 10^7\ \mathrm{yr})\left(\frac{M}{4\eta\,M_\odot}\right)^{1/5}.
\label{tN}
\ee

For this sequence to be possible within the age of the universe, $t_{\mathrm{total}}<t_0$, in the best case scenario of equal masses for each binary at each stage, so that $4\eta = 1$, one needs
\bb
M < M_{\mathrm{max}} \approx 1.4\times 10^{14} M_\odot.
\label{Mma}
\ee
This is what this procedure, with the timescale $T$ obtained from Eq.\ (43) of \cite{Milosavljevic:2001vi} and defined by Eq.\ (\ref{T}) above, gives for a crude approximation for this conjectured way of getting an upper mass for a black hole that could form within the 14 Gyr age of the present universe.

Putting the pieces together, and using $T \approx 210\, \mathrm{Gyr} \approx 15\, t_0$, one gets following formula for the maximum mass after a sequence of many equal-mass binary coalescences taking up the full age of the universe $t_0 \approx 14\, \mathrm{Gyr}$:
\bb
M_{\mathrm{max}} = \frac{16}{5}(1-2^{-1/5})^5
\left(\frac{125}{2\pi^2(5+\sqrt{5})}\right)^{5/2} \frac{c^3}{G} \frac{t_0^5}{T^4}
\approx \frac{c^3 t_0^5}{12\,000\,G\, T^4} \approx 1.4\times 10^{14} M_\odot.
\label{Mmax}
\ee

One main implication of this result is that the canonical frequency dependence, $f^{-2/3}$, for the spectrum of the gravitational-wave strain from sources that have already coalesced \cite{Phinney:2001di}, will not rise so fast as the frequency $f$ is lowered below that emitted by the maximum-mass coalesced black hole binaries.  For any binary evolution obeying Eq.\ (\ref{vdotgr}) above, for $x \ll 1$ the binary will lose energy mainly by gravitational interactions with the surroundings (e.g., stars), rather than by radiating gravitational waves, and only for $x \geq 1$ will gravitational waves dominate.  Assuming quasi-circular orbits (relatively slow inspiral in the nonrelativistic regime), the frequency $f$ of gravitational waves (twice that of the orbits) is
\bb
f = \frac{1}{\pi}\frac{c^3}{GM}\left(\frac{v}{c}\right)^3.
\label{f}
\ee

During inspiral, the dominant gravitational wave strain will be produced when
\bb
x \equiv \left(\frac{16\,(4\eta)\,c^3T}{5\,GM}\right)^{1/5}
\left(\frac{v}{c}\right)^{2} 
\equiv \left(\frac{4\eta\, M_*}{M}\right)^{1/5}
\left(\frac{v}{c}\right)^2 \sim 1.
\label{xp}
\ee
Taking the best case scenario of equal binary masses, $M_1=M_2$, so that in this case $4\eta \equiv 4M_1M_2/(M_1+M_2)^2 = 1$, and defining
\bb
K \equiv \left(\frac{M_*}{M_{\mathrm{max}}}\right)^{1/5} \approx 125,
\label{K}
\ee
one gets
\bb
x = K\left(\frac{M_{\mathrm{max}}}{M}\right)^{1/5}
\left(\frac{v}{c}\right)^2 = 1
\label{xK}
\ee
at
\bb
\left(\frac{v}{c}\right)^2 = \frac{1}{K}\left(\frac{M}{M_{\mathrm{max}}}\right)^{1/5},
\label{beta}
\ee
giving
\bb
f = \frac{c^3 K^{-3/2}}{\pi GM_{\mathrm{max}}}\left(\frac{M_{\mathrm{max}}}{M}\right)^{7/10}
\approx 0.33\left(\frac{M_{\mathrm{max}}}{M}\right)^{7/10}\,\mathrm{picohertz},
\label{fm}
\ee
or a gravitational wave period $P$ of
\bb
P \approx 96\,000 \left(\frac{M}{M_{\mathrm{max}}}\right)^{7/10}\,\mathrm{yr}.
\label{P}
\ee
Below this frequency $f$, or longer than this period $P$, for supermassive black hole binary coalescence within the present age of the universe, the gravitational wave strain should no longer increase as the frequency decreases.

This work was supported in part by the Natural Sciences and Engineering Research Council of Canada. 

%\newpage

\end{document}